\documentstyle[preprint,eqsecnum,aps,epsf]{revtex}
\input{psfig}
\begin{document}
\newcommand \beq { \begin{eqnarray} }
\newcommand \eeq { \end{eqnarray} }
\vspace{1cm}
\begin{center}
{\large\bf Photon-Photon and Pomeron-Pomeron Processes in Peripheral Heavy
Ion Collisions}\\[.4cm]       
C.~G.~Rold\~ao~\footnotemark 
\footnotetext{e-mail : roldao@ift.unesp.br} \hspace{2mm} and 
A.~A.~Natale~\footnotemark\\[.2cm]
\footnotetext{e-mail : natale@ift.unesp.br}
Instituto de F\'{\i}sica Te\'orica,   
Universidade Estadual Paulista\\
Rua Pamplona, 145, 01405-900, S\~ao Paulo, SP, Brazil
\end{center}
\thispagestyle{empty}
\vspace{1cm}

\begin{abstract}						            
We estimate the cross sections for the production of resonances, pion pairs
and a central cluster of hadrons in peripheral  heavy-ion collisions through
two-photon and double-pomeron exchange, at  energies that will be available at
RHIC and LHC. The effect of the impact parameter in the diffractive reactions
is introduced, and imposing the condition for realistic peripheral collisions 
we verify that in the case of very heavy ions the pomeron-pomeron contribution
is indeed smaller than the electromagnetic one. However, they give a
non-negligible background in the collision of light ions. This diffractive 
background will be more important at RHIC than at LHC.
\end{abstract}

\draft


\newpage                      
\section{Introduction}

\noindent

Collisions at relativistic heavy ion colliders like the Relativistic
Heavy Ion Collider RHIC/Brookhaven and the Large Hadron Collider LHC/CERN
(operating in its heavy ion mode) are mainly devoted to the search of
the Quark Gluon Plasma. However, peripheral heavy ion collisions also
open up a broad area of studies as advocated by Baur and 
collaborators~\cite{baur,baur2}. Examples are the possible discovery of an
intermediate-mass Higgs boson~\cite{papa,cahn} or beyond standard model
physics~\cite{ns} using peripheral ion collisions, which have been discussed
at length in the literature. More promissing than these may be the study of
hadronic physics, which will appear quite similarly to the two-photon
hadronic physics at $e^+e^-$ machines with the advantage of a huge photon
luminosity peaked at small energies~\cite{baur,baur2,klein}. Due to this large
photon luminosity it will become possible to discover resonances
that couple weakly to the photons~\cite{natale}.

Double-pomeron exchange will also occurs in peripheral heavy ion collisions
and their contribution is similar to the two-photons one as discussed by
Baur~\cite{baur} and Klein~\cite{klein}. A detailed calculation performed by
M\"uller and Schramm of Higgs boson production have shown that the diffractive
contribution is much smaller than the electromagnetic one~\cite{muller}. We
can easily understand this result remembering that the coupling between the
Higgs boson and the pomerons is intermediated by quarks, and according to
the pomeron model of Donnachie and Landshoff~\cite{land1} when in the
vertex pomeron-quark-quark any of the quark legs goes far ``off-shell" the
coupling with the pomeron decreases. Therefore, we do not need to worry
about the pomeron-pomeron contribution in peripheral heavy ion collisions
when heavy (or far ``off-shell") quarks are present. However, this is not
what happens in the case of light resonances~\cite{natale}, where double
diffraction were claimed to be as important as photon initiated processes. In
particular, 
Engel {\it et al.}~\cite{engel} have shown that at the LHC the diffractive
production of hadrons may be a background for the photonic one. 

In Ref.~\cite{baur} it was remarked that the effect of removing ``central
collisions" should also be performed in the double-pomeron calculation,
implying in a considerable reduction of the background calculated in
Ref.\cite{engel}. This claim is the same presented by Baur~\cite{baur2} and
Cahn and Jackson~\cite{cahn} in the case of early calculations of peripheral
heavy ion collisions. Roughly speaking we must enforce that the minimum impact
parameter ($b_{min}$) should be larger than ($R_1 + R_2$), where $R_i$ is the
nuclear radius of the ion ``$i$", in order to have both ions coming out intact
after the interaction.

In this work we will compute the production of resonances, pion-pairs
and a hadron cluster with invariant mass $M_X$ through photon-photon
and pomeron-pomeron fusion in peripheral heavy ion collisions at the energies
of RHIC and LHC. We will take into account the effect of the impact
parameter as discussed in the previous paragraph for photons as well
as for pomerons. We also compare this approach to cut the central
collisions with the use of an absorption factor in the Glauber
approximation.
The inclusion of pion-pairs production is important
because they certainly will be studied at these colliders, and they
also represent a background for glueball (and other hadrons) detection.
The pomeron physics within the ion will be described by the 
Donnachie and Landshoff model~\cite{land1,land2}. We will focus on
the values of the cross sections that shall be measured in the already
quoted ion colliders, and point out when pomeron-pomeron processes
can be considered competitive or not with photon-photon collisions.
The arrangement of our paper is the following: Section 2 contains
a discussion of the photons and pomerons distributions in the nuclei.
In Sect. 3 we introduce the cross section for the elementary processes.
Finally, Sect. 4 contains the results and conclusions.

\section{Photons and pomerons distribution functions}
                          
\subsection{Photons in the nuclei}

${}$
\vskip -0.5cm

The photon distribution in the nucleus can be described using the 
equivalent-photon or Weizs\"{a}cker-Williams approximation in the impact
parameter space. Denoting by $F(x)dx$ the number of 
photons carrying a fraction between $x$ and $x+dx$ of the 
total momentum of a nucleus of charge $Ze$, we can define 
the two-photon luminosity through
\beq
\frac{dL}{d\tau} = \int ^1 _\tau \frac{dx}{x} F(x) F(\tau/x),
\eeq

\noindent
where $\tau = {\hat s}/s$, $\hat s$ is the square of the center of mass
(c.m.s.) system energy of the two photons and $s$ of the ion-ion system. The
total cross section $ ZZ \rightarrow ZZ \gamma \gamma \rightarrow ZZ X$, where
$X$ is the particle produced within the rapidity gap, is
\beq
\sigma (s) = \int d\tau \frac{dL}{d\tau} \hat \sigma(\hat s),
\label{sigfoton}
\eeq

\noindent
where $ \hat \sigma(\hat s)$ is the cross-section of the subprocess
$\gamma \gamma \rightarrow X$.

There remains only to determine $F(x)$. In the literature there are
several approaches for doing so, and we choose the conservative and
more realistic photon distribution of Ref.\cite{cahn}. Cahn and Jackson~\cite{cahn}, using a prescription
proposed by Baur~\cite{baur}, obtained a photon
distribution which is not factorizable. However, they were able to give a fit
for the differential luminosity which is quite useful in practical
calculations: 					
\begin{equation}							    
\frac{dL}{d\tau}=\left(\frac{Z^2 \alpha}{\pi}\right)^2 \frac{16}{3\tau} 
\xi (z),
\label{e3}		
\end{equation}								    
where $z=2MR\sqrt{\tau}$, $M$ is the nucleus mass, $R$ its radius and 
$\xi(z)$ is given by 							    
\begin{equation}							    
\xi(z)=\sum_{i=1}^{3} A_{i} e^{-b_{i}z},   				    
\label{e4}								    
\end{equation}
which is a fit resulting from the numerical integration of the photon 
distribution, accurate to $2\% $ or better for $0.05<z<5.0$, and where 
$A_{1}=1.909$, $A_{2}=12.35$, $A_{3}=46.28$, $b_{1}=2.566$, 
$b_{2}=4.948$, and $b_{3}=15.21$. For $z<0.05$ we use the expression (see 
Ref.~\cite{cahn}) 
\begin{equation}			                                    
\frac{dL}{d\tau}=\left(\frac{Z^2 \alpha}{\pi}\right)^2 
\frac{16}{3\tau}\left(\ln{(\frac{1.234}{z})}\right)^3 .
\label{e5}								    
\end{equation}								    
The condition for realistic peripheral collisions ($b_{min} > R_1 + R_2$) is
present in the photon distributions showed above, and the applications
of Sect. 4 are straightforward once we determine the cross sections for the
elementary processes.
 
\noindent

\subsection{Pomerons in the nuclei}

${}$
\vskip -0.5cm

In the case where the intermediary particles exchanged in the nucleus-nucleus 
collisions are pomerons instead of photons, we can follow closely the
work of M\"{u}ller and Schramm~\cite{muller} and make a
generalization of the equivalent photon approximation method to this new
situation. So the cross section for particle production via two pomerons
exchange can be written as  %
\beq 
\sigma_{AA}^{PP} = \int dx_1 dx_2 f_P(x_1)f_P(x_2) \sigma_{PP}(s_{PP}),
\label{sechoque}
\eeq

\noindent
where $f_P(x)$ is the distribution function that describe the probability for
finding a pomeron in the nucleus with energy fraction $x$ and
$\sigma_{PP}(s_{PP})$ is the subprocess cross section 
with energy squared $s_{PP}$. In the case of inclusive particle production
we use the form given by
Donnachie and Landshoff \cite{land3} 
\beq 
f_P(x) = \frac{1}{4 \pi ^2 x} \int ^{-(xM)^2} _{-\infty} dt \,
 | \beta_{AP}(t)|^2 \, |D_P(t;s^\prime)|^2, 
\label{in} 
\eeq

\noindent
where $D_P(t;s^\prime)$ is the pomeron propagator\cite{land2} 
\beq
D_P(t;s) = \frac{(s/m^2)^{\alpha_P(t)-1}}{\sin ( \frac{1}{2} \pi \alpha _P(t))}
\exp{\left( - \frac{1}{2} i \pi \alpha _P(t) \right)}, \nonumber 
\eeq

\noindent
with $s$ the total squared c.m. energy.
The Regge trajectory obeyed by the pomeron is 
$\alpha _P(t) = 1 + \varepsilon + \alpha ^\prime _P t$, where $\varepsilon =
0.085$, $ \alpha ^\prime _P =0.25$ GeV$^{-2}$ and $t$ is a small exchanged
four-momentum square, $t= k^2 <<1$, so the pomeron behaves like
a spin-one boson. The term in the denominator of the pomeron propagator,
$ [\sin (\frac{1}{2} \pi \alpha _P(t))]^{-1}$, is the signature factor that
express the different properties of the pomeron under C and P conjugation. At
very high c.m. energy this factor falls very rapidly with ${\mathbf{k}} ^2$,
whose exponential slope is given by $\alpha ^\prime _P \ln(s/m^2)$, $m$ is the
proton mass, and it is possible to neglect this ${\mathbf{k}}^2$ dependence, 
\beq 
\sin \frac{1}{2} \pi (1+ \varepsilon - \alpha ^\prime _P {\mathbf{k}}^2)
\approx \cos (\frac{1}{2} \pi \varepsilon ) \approx 1. \nonumber 
\eeq

\noindent
If we define the pomeron range parameter $r_0$ as
\beq
r_0 ^2 = \alpha ^\prime _P \ln (s/m^2),
\label{r0}
\eeq

\noindent
the pomeron propagator can be written as
\beq
|D_P(t=-{\mathbf{k}}^2;s)| = (s/m^2)^\varepsilon e^{-r_0^2 {\mathbf{k}}^2}.
\label{pomprop}
\eeq

\noindent
Since we are interested in the spatial distribution of the virtual quanta in
the nuclear rest frame we are using $t=-{\mathbf{k}}^2$.

The nucleus-pomeron coupling has the form \cite{land3}
\beq
\beta_{AP}(t)= 3 A\beta_0 F_A(-t),
\nonumber
\eeq

\noindent
where $\beta _0 = 1.8 $ GeV$^{-1}$ is the pomeron-quark coupling, $A$ is the
atomic number of the colliding nucleus, and 
$F_A(-t)$ is the nuclear form factor for which is usually
assumed a Gaussian expression (see, e.g., Drees et al. in \cite{papa})
\beq
F_A(-t)= e^{t/2Q_0^2},
\label{fatf}
\eeq

\noindent
where $Q_0=60$ MeV.

Performing the $t$ integration of the distribution function in Eq.(\ref{in}) we obtain
\beq
f_P(x) &=& \frac{(3 A \beta_0)^2}{(2 \pi)^2 x} 
\left( \frac{s^\prime}{m^2} \right)^{2 \varepsilon} \int ^{-(xM)^2}
_{-\infty} dt \,  e^{t/Q_0^2} \nonumber \\
&=& \frac{(3 A \beta_0 Q_0)^2}{(2 \pi)^2x} 
\left( \frac{s^\prime}{m^2} \right)^{2 \varepsilon} \exp \left[-\left(\frac{xM}{Q_0}\right)^2 \right]. 
\nonumber
\eeq

The total cross section for a inclusive particle production is obtained with
the above distribution and also with the expression for the subprocess $PP
\rightarrow X$ as prescribed in Eq.(\ref{sechoque}). However,
in Eq.(\ref{sechoque}) the cases where the two nuclei overlap are not
excluded. To enforce the realistic condition of a peripheral collision  
it is necessary to perform the calculation taking into account the impact
parameter dependence, $b$. It is straightforward to verify that in the
collision of two identical nuclei the total cross section of Eq.(\ref{sechoque})
is modified to~\cite{muller}
\beq
\frac{d^2 \sigma ^{PP \rightarrow X}_{AA}}{d^2 b} = 
\frac{Q^{\prime 2}}{2 \pi} \, e^{-Q^{\prime 2} b^2/2} \, \sigma ^{PP}_{AA},
\label{dsin}
\eeq

\noindent
where $(Q^{\prime })^{-2} = (Q_0)^{-2} + 2 r_0 ^2$. The total
cross section for inclusive processes is obtained after integration of
Eq.(\ref{dsin}) with the condition $b_{min} > 2R$ in the case of identical
ions.

For exclusive particle production the determination of the pomeron distribution
function in the nuclei is slightly modified,
because in this case it is necessary some specific
assumption about the pomeron internal structure \cite{schafer}. Following
Ref.\cite{muller} the distribution function of pomerons is
\beq
f_P(x) = \frac{(3 A \beta_0)^2}{(2 \pi)^2 x} \int ^{-(xM)^2} _{-\infty} dt \, 
(-t-x^2M^2) \,
F_A(-t)^2 \, |D(t)|^2,
\label{ex}
\eeq
and the cross section for a resonance production as a function of
the impact parameter is~\cite{schramm}   
\beq
\frac{d^2 \sigma ^{PP\rightarrow R}_{AA}}{d^2 b} &=& 
2 \pi \left( \frac{ 3 A \beta_0}{2 \pi ^2} \right)^4 \int \frac{dx_1}{x_1} \,
 \frac{dx_2}{x_2} \,  Q_1^4 \,  Q_2^4  \, \tilde{Q}^2 \, e^{-x_1 ^2 M^2/Q_1^2}
\, e^{-x_2 ^2 M^2/Q_2^2}  \nonumber \\ &\times&\left( \frac{x_1 x_2 s^2}{m^4}
\right)^{2 \varepsilon} \,\sigma ^{PP\rightarrow R}_{AA}(x_1 x_2 s) \, b^2 \,
\tilde{Q}^2 \, e^{-b^2 \tilde{Q}^2/2}, 
\nonumber
\eeq

\noindent
with $\sigma ^{PP\rightarrow R}_{AA}(x_1 x_2 s)$ indicating the subprocess
cross section (double pomeron fusion producing a resonance), and
where
\beq
\tilde{Q}^{-2} = \frac{1}{2}(Q_1^{-2} + Q_2^{-2}),
\nonumber
\eeq

\noindent
with $Q_i^{-2} \equiv Q_0^{-2} + 2 r_0 ^2$ for idential ions. In the
calculations we are going to perform we noticed that the
aproximation $Q^{-2} _i \approx Q_0^{-2}$ is quite reasonable, because for the
energies that we shall consider the
pomeron range parameter (Eq.(\ref{r0})) is smaller than the width of the
Gaussian form factor and consequently $\tilde{Q}^2 \approx Q^2_0$. Therefore,
we obtain the final expression  
\beq
\frac{d^2 \sigma ^{PP\rightarrow R}_{AA}}{d^2 b} &=& 
2 \pi \left( \frac{ 3 A \beta_0 Q_0^2}{2 \pi ^2}  \right)^4 \int
\frac{dx_1}{x_1} \,  \frac{dx_2}{x_2}   \,  e^{-x_1 ^2 M^2/Q_0^2} \, e^{-x_2
^2 M^2/Q_0^2}  \nonumber \\
 &\times&\left( \frac{x_1 x_2 s^2}{m^4} \right)^{2
\varepsilon} \,\sigma ^{PP\rightarrow R}_{AA}(x_1 x_2 s) \, b^2 \, Q_0^4 \,
e^{-b^2 Q_0^2/2}. 
\label{dsex2} 
\eeq
As discussed previously, to enforce the condition of peripheral collisions
we integrate Eq.(\ref{dsex2}) with the condition $b_{min} > 2R$.

Another way of to exclude events due to inelastic central collisions is
through the introduction of an absortion factor computed in the Glauber
aproximation \cite{glauber}. This factor modifies the cross section
in the following form 
\beq \frac{d \sigma^{gl}_{AA}}{d^2 b} &=& 
\frac{d \sigma_{AA}^{{ PP} \rightarrow R}}{d^2 b} \, \exp \left[ -A^2 b
\sigma_0 \int \frac{dQ^2}{(2 \pi )^2} \, F_A^2(Q^2) \, e^{iQb}\right]
\nonumber \\ 
&=& \frac{d \sigma_{AA}^{{ PP} \rightarrow R}}{d^2 b} \, \exp
\left[ -A^2 b \sigma_0 \,\frac{Q_0^2}{4 \pi} \, e^{-Q_0^2 b^2/4} \right],
\label{gla} 
\eeq

\noindent
where $\sigma _0$ is the nucleon-nucleon total cross section, whose 
value for the different energy domains that we shall consider is obtained
directly from the fit of Ref. \cite{caso} 
\beq
\sigma_0= X s^\epsilon + Y_1 s^{-\eta _1} +  Y_2 s^{-\eta _2},
\nonumber
\eeq

\noindent
with $X = 18.256$, $Y_1 = 60.19$, $Y_2 = 33.43$, $\epsilon = 0.34$, $\eta _1 = 0.34$, $\eta _2 = 0.55$,
$F_A(Q^2) = e^{-Q^2/2Q^2_0}$ and we exemplified Eq.(\ref{gla}) for the case of
resonance production, {\it i.e.},  $\sigma_{AA}^{ PP \rightarrow R}$ is
the total cross section for the resonance production to be discussed in the
next section. The integration in Eq.(\ref{gla}) is over all impact parameter
space and in the last section we discuss the differences between the two
approaches showed above for removing central collisions.

\section{Subprocesses initiated by photons and pomerons}

\subsection{Resonances}

${}$
\vskip -0.5 cm

The main motivation to study resonance production in peripheral heavy
ion collisions is that the high photon luminosity will allow us to
observe resonances that couple very weakly to photons. 
The simplicity of this calculation also enable us to test the
methods for removing central collisions, as well as to check up
to which degree the double pomeron exchange is or not a background
for the two photon physics.

To estimate the production of single spin-zero resonances, we note that
these states can be formed by photon-photon fusion with a coupling strength
that is measured by their photonic width
\beq
\hat{\sigma}_{\gamma \gamma \rightarrow R} = \frac{8 \pi ^2}{M_R \hat{s}} \,
\Gamma  _{R \rightarrow \gamma \gamma} \,  \delta \left( \tau -
\frac{M^2_R}{\hat{s}} \right),
\eeq

\noindent
where $M_R$ is the ressonance mass and 
$\Gamma  _{R \rightarrow \gamma \gamma}$ its decay width in two photons.
Using this expression into Eq.(\ref{sigfoton}) we obtain the total cross
section for the production of pseudo-scalar mesons. 

To compute the cross section of the 
subprocess $PP \rightarrow R$ we can use the pomeron model of 
Donnachie and Landshoff~\cite{land2}. In this model it is
assumed that the pomeron couples to the quarks like a
isoscalar photon \cite{land2}. This means that the cross sections of $PP \rightarrow X$ 
subprocesses can be obtained from suitable modifications on the cross-section
for $\gamma \gamma \rightarrow X$. Another aspect to be considered
is that the pomeron-quark-quark vertex is not point-like, and 
when either or both of the two quark legs in this vertex goes far off
shell the coupling is known to decrease. 
So the quark-pomeron coupling $\beta_0$  must be replaced by  
\beq
\tilde{\beta}_0(q^2) = \frac{\beta_0 \, \mu _0^2}{\mu^2_0 + Q^2},
\label{btilde} 
\eeq

\noindent
where $\mu _0^2=1.2$ GeV$^2$ is a mass scale characteristic of the pomeron, in
the case of resonance production 
$Q= M_R/2$ measures how far one of the quark legs is off mass shell and $M_R$
is the resonance mass. 
Therefore, the process $PP \rightarrow R$
is totally similar to the one initiated by photons unless from an
appropriate change of factors. The cross section we are looking for is
obtained changing
the fine-structure constant $\alpha $ by $9 \tilde{\beta}/16 \pi ^2$, where 
$\tilde{\beta}$ is giving by Eq.(\ref{btilde}) and $9=3^2$ is a color factor,
leading to
\beq 
\sigma ^R_{ PP} = \frac{9}{2} \, \frac{\tilde{\beta} ^4}{\alpha ^2} \,
\frac{ \Gamma (R \rightarrow \gamma \gamma)}{M_R} \delta(x_1 x_2 s - M_R^2).
\nonumber 
\eeq 

\noindent
Using this expression in Eq.(\ref{dsex2}) the total
cross section is equal to
\beq
\sigma ^{ PP \rightarrow R}_{AA} &=&
\frac{9 \pi}{8} \frac{(\tilde{\beta}Q_0)^4}{\alpha ^2} 
\left( \frac{3 A\beta_0 Q_0}{2 \pi} \right)^4 \frac{\Gamma(R \rightarrow
\gamma \gamma )}{M_R} \left( \frac{M_R^2 s}{m^4} \right)^{2 \epsilon}
\frac{Q^4_0}{M_R^2} 
\nonumber \\ 
&\times & \int \frac{dx}{x} \exp \left[
\left( -\frac{M_R^2 M}{s Q_0 x} \right)^2 -\frac{(xM)^2}{Q_0^2} \right] \int
^\infty _{b_{min}} db 2 \pi b^3 e^{-Q_0^2b^2/2}, 
\nonumber \\ 
\label{sigmaex}
\eeq

\noindent
where $b_{min} = 2R$. 

\subsection{Pion pair production}

${}$
\vskip -0.5cm

The continuous production of pion pairs ($\pi ^+ \pi ^-$) is also an
interesting signal to be observed in peripheral heavy ion collisions,
mostly because they are a background for glueball and other resonances
decays. Here we discuss the subprocess cross sections for two photons,
$ZZ \rightarrow \gamma \gamma 
\rightarrow ZZ \pi ^+ \pi ^-$,
and two pomerons exchange $ZZ \rightarrow PP \rightarrow ZZ \pi ^+ 
\pi^-$.

The cross section for pion pair production by two photons can be calculated
approximately by using a low energy theorem derived from 
partially-conserved-axial-vector-current hypothesis and current 
algebra and is equal to~\cite{ter} 
\beq \sigma(\gamma \gamma
\rightarrow \pi^+ \pi^-) \cong \frac{2 \pi \alpha ^2}{s} \left( 1-
\frac{4 m_\pi ^2}{s} \right)^{(1/2)} \left[ \frac{ m^4_V}{ \left(
\frac{1}{2}s + m^2_V \right) \left( \frac{1}{4}s + m^2_V \right)
} \right]^2,
\label{sigpp}
\eeq

\noindent
where $m_{\pi}$ is the pion mass an $s$ its squared energy, $m_V$ 
is a free parameter, whose value that provides the best fit to the
experimental data is $m_V \cong 1.4$ GeV. This expression shows a nice
agreement with the experimental data~\cite{cleo}. For large values of $s$
it deviates from the Brodsky and Lepage formula~\cite{brod}. However, since
most of the photon distribution is concentrated in the small $x$ region,
{\it i.e.}, the photons carry a small fraction of the momentum
of the incoming ion, the difference is negligible.

Using Eqs. (\ref{sigpp}) and (\ref{sigfoton}) we obtain
\beq
\sigma (s) = \frac{2 \pi \alpha ^2}{s} 
\int ^1 _{\tau_{min}} \frac{d\tau }{\tau } \left( 1- \frac{4 m_\pi ^2}{s
\tau} \right)^{(1/2)} \left[ \frac{ m^4_V}{ \left( \frac{1}{2}s \tau + m^2_V
\right) \left( \frac{1}{4}s \tau  + m^2_V \right) } \right]^2 \,
\frac{dL}{d\tau}. 
\nonumber 
\eeq

In the case of double pomeron exchange producing a pion pair we use once
again the Donnachie and Landshoff model for the pomeron, obtaining
the cross section for $ PP \rightarrow \pi ^+ \pi^-$ from the photonic one
changing $\alpha ^2 \rightarrow 9 \tilde{\beta}^4_0/16 \pi^2$ in $\sigma
(\gamma \gamma \rightarrow \pi ^+ \pi^-)$, and the resulting expression
replaces $\sigma ^{PP \rightarrow R}_{AA}(x_1 x_2 s)$ in Eq.(\ref{dsex2}). The total cross
section appears after we perform the integration in the parameter space
representation of the following equation 
\beq 
\frac{d^2 \sigma ^{PP \rightarrow \pi ^+ \pi ^-}_{AA}}{db^2}
&=& \left( \frac{\pi ^2}{8} \right)  \frac{9}{4}  \frac{(\tilde{\beta}_0
Q_0)^4}{s} \left( \frac{3 A \beta _0 Q_0}{2 \pi^2} \right)^4 \int
\frac{dx_1}{x_1^2} \, \frac{dx_2}{x_2^2} e^{-(x_1 M)^2/Q_0^2} e^{-(x_2
M)^2/Q_0^2} 
\nonumber \\ 
&\times& \left( \frac{x_1 x_2 s^2}{m^4} \right)^{2
\varepsilon} \left( 1 - \frac{4 m^2_\pi}{x_1 x_2 s} \right)^{1/2} \left[
\frac{m_V^4}{\left( \frac{x_1 x_2 s}{2} + m_V^2 \right) \left( \frac{x_1 x_2
s}{4} + m_V^2 \right)} \right] ^2 
\nonumber \\ 
&\times& \int ^\infty
_{b_{min}}db 2 \pi Q_0 ^4 b^3 e^{-b^2 Q_0^2/2}. 
\nonumber 
\eeq

\subsection{Multiple particle production}

${}$
\vskip -0.5 cm

The elementary cross section for multiple-particle production via two photons
fusion can be described by the parametrization \cite{l3}
\beq
 \sigma _{\gamma \gamma \rightarrow hadrons}= C_1 \left( \frac{s}{s_0} \right)
^\epsilon +  C_2  \left( \frac{s}{s_0} \right) ^{-\eta},
\label{l3}
\eeq

\noindent
where $C_1 = 173$ nbarn, $C_2 = 519$ nbarn, $s_0 = 1$ GeV$^2$, $\epsilon = 0.079$ 
and $\eta = 0.4678$. The total cross section comes out from 
Eq.(\ref{sigfoton}).

Within the Donnachie and Landshoff model it is straightforward to see that
with the above parametrization the differential cross section to produce
a cluster of particles with mass $M_X$ through double pomeron exchange is
\beq \frac{d \sigma}{dM_X} &=& \frac{(3 A \beta_0  \tilde{\beta}_0
\mu_0)^4}{(2 \pi )^4 R^4_N (16 \pi ^2 \alpha ^2)} \, \frac{1}{2 M_X}  \int
\frac{ds^\prime}{s^\prime} \, \left[ C_1 \left( \frac{s^\prime}{s_0} \right)
^\epsilon + C_2 \left( \frac{s^\prime}{s_0} \right)^{-\eta} \right] \, 
\nonumber
\\ &\times& \exp \left[ - \left( \frac{s^\prime M R_N}{s} \right)^2 - \left(
\frac{M^2_X M R_N}{s^\prime} \right)^2 \right] \int ^\infty _{b_{min}} db \, b
\, \frac{e^{-b^2/2 R_N^2}}{R_N^2},
\nonumber 
\eeq

\noindent
to obtain this expression we used the pomeron distribution function
in the nucleus for inclusive process (Eq.(\ref{in})).

To be possible a comparison with the work of Engel {\sl et al.}~\cite{engel}, we
also make use of the Ter-Martirosyan \cite{mart} model for diffractive
multiparticle production. In this model
the subprocess $PP \rightarrow hadrons$ is characterized by the
cross section
\beq \sigma_{ PP}^{tot}(\ln(M^2_X/m^2),t_1,t_2) \approx 8 \pi r(t_1) r(t_2),
\label{sigmar}
\eeq
which is a function of the triple-pomeron vertex
$r(t)$, where t is the exchanged momentum. Using the value of $r(0)$ from 
Ref. \cite{cool}, $\sigma_{\cal PP}^{tot} = 8 \pi r^2(0) \approx 140$ $\mu$ barn.
Note that we have clear differences between the approaches described
above. Eq.(\ref{l3}) is a parametrization valid for a wide range of momenta,
and with this one we naively apply the model of Ref. \cite{land1} to compute
the total cross section for multiparticle production. On the other hand
Eq.(\ref{sigmar}) is obtained in another specific model and it is not
expected to be valid for the same range of energies as Eq.(\ref{l3}). This
difference is going to be discussed in the last section.

Streng \cite{streng} applied the model of Ref.~\cite{mart}
for proton-proton collisions where the initial protons are 
scattered almost elastically, 
emerging with a very large fraction of the initial energy, 
\beq 
|x_1|,|x_2| \geq c, \, c \geq 0.9. \nonumber 
\eeq

\noindent
The double pomeron exchange produce a particle cluster within a large
rapidity gap and with a mass of the order
\beq
M^2_X \approx s(1-|x_1|) (1-|x_2|),
\label{mx}
\eeq

\noindent
where $s$ is the reaction energy squared.
As the scattering is almost elastic, {\it i.e.}, the emerging beam has 
approximately the same energy as the incident one, the following
kinematical boundaries can be introduced
\beq
M_0 \leq M_X \leq (1 - c) \sqrt{s},
\nonumber \\
\frac{M^2_X}{(1-c)} \leq s_1 \leq (1-c)s
\label{lim},
\eeq

\noindent
where $M_0 = 2$ GeV and $c = 0.9$. These limits have been translated for the
case of heavy ions by Engel {\sl et al.}~\cite{engel}
and we will proceed as them. If we consider Eq.(\ref{sigmar}),
dress it with the pomeron distribution functions within the
nuclei, and subtract the central collisions considering the
absortion factor computed in the Glauber
aproximation \cite{glauber} we reproduce the results of Ref.~\cite{engel}.

\section{Results and Conclusions}

${}$
\vskip -0.5 cm

Peripheral collisions at relativistic heavy ion colliders provide an arena
for interesting studies of hadronic physics. Resonances coupling weakly to
photons can be studied due to the large photon luminosity, the continuous
production of pion pairs will be observed not only as a reaction of interest
as well as a possible background for some resonance decays. A hadron cluster
produced within a large rapidity gap will give information about photon-photon
and double pomeron exchange. In this work we estimate the cross section
for these processes. One of the main points is to verify if the double
pomeron exchange is or not a background for the purely electromagnetic
process. We discussed double pomeron exchange according to the
Donnachie and Landshoff~\cite{land2} model and calculated the cross sections
in the impact parameter space. The condition for a realistic peripheral
collision is imposed integrating the cross section with $ b_{min} > 2R$
in the case of two identical ions with radius $R$.

We considered the production of pseudoscalars resonances in
the collision of $^{238}$U for energies available 
at RHIC ($\sqrt{s} = 200 $ GeV/nucleon), and collisions of $^{206}$Pb at
energies available in LHC ($\sqrt{s} = 6.300$ GeV/nucleon). Our results 
are shown in Table 1. Contrarily to the result of Ref.~\cite{natale}
the double pomeron exchange is not important when the cut in the
impact parameter is introduced. For a realistic peripheral collision
in the case of resonance production the pomeron-pomeron process is
at least two orders of magnitude below the photon-photon one.
Note in Table 1 that the rate of diffractive resonance production decreases
with the increase of the meson mass. The main reason for this behavior lies
in the fast decrease of the pomeron-quark coupling as shown in
Eq.(\ref{btilde}).

Note that the results of this table assume $100 \% $ of efficiency in
tagging the peripheral collision, even if we consider a small 
efficiency we recall that the cross section for light resonances
imply in approximately billions of events/yr which easily survive the
cuts for the background separation proposed by Nystrand and 
Klein (see the last paper of Ref.~\cite{klein}). One of the most important
cuts to separate inelastic nuclear reactions, associated with
grazing collisions, is the small multiplicity of the final state, and
this is exactly what we may expect in the final state of the particles
discussed in Table 1. 

The decays of $\pi^0$, $\eta$, etc... will be
dominated by two (or three) body decays in the central region of rapidity,
and easily separated from the larger multiplicity common to inelastic
collisions. It is interesting that in the case of $\pi^0$ and $\eta$
production we may focus on the $2 \gamma$ decay, and even if it is possible
to separate the background from inelastic nuclear reactions, we still
have the background of the photon-photon scattering through the QED box 
diagram producing the same final state. The box 
diagram will be dominated by light quarks, electron and muon, and for 
these we can use the asymptotic expression of $\gamma 
\gamma$ scattering ($\sigma(s) \sim	20/s$). Integrating this expression 
in a bin of energy (proportional  to the resonance partial width into
two-photons) centered at the mass of the resonance, we obtain a  cross-section
smaller than the resonant one with subsequent decay into  two-photons. We do
not considered the interference between the box and resonant diagram because
on resonance the two processes are out of phase. It is opportune to mention
that the decay products will fill the central region of rapidity, which is
also one of the conditions proposed in Ref.~\cite{klein} to isolate the
peripheral collisions.

As discussed in Sect. 2 we have two different ways to enforce a realistic
peripheral heavy ion collision. One is a geometrical cut in the impact
parameter space where $b_{min} > 2R$ is imposed, the other is through
the introduction of the absorption factor in the Glauber approximation
as given by Eq.(\ref{gla}).
In Table 2 we compare the ratios between the total cross sections for
diffractive resonance production computed with Eq.(\ref{gla}) and the
one with the cut on the impact parameter (given by Eq.(\ref{sigmaex})),
in
the collision of $^{238}$U for energies available 
at RHIC ($\sqrt{s} = 200 $ GeV/nucleon), and collisions of $^{206}$Pb at
energies available in LHC ($\sqrt{s} = 6.300$ GeV/nucleon).

The values of Table 2 show that the geometrical cut is less
restrictive than the one given by the Glauber absorption factor. However,
which one is more realistic also depends on the energy and on the ion
that we are considering. In Table 3 we present the cross section for $\pi ^0$ 
production for different ions and at different energies.
From Eq.(\ref{gla}) we notice that small variations in $\sigma _0$ (the
nucleon-nucleon total cross section) are also promptly transmitted to the total
cross section, and modify the ratios between the different methods to exclude
inelastic collisions. Table 3 shows that the difference between the methods
also become less important for light ions, but the most striking fact in
this table is that for light ions double pomeron exchange starts becoming
a background for photon-photon processes! According to Table 3 for $^{28}$Si
the diffractive $\pi ^0$ production is a factor of 2 down the
electromagnetic one (assuming the geometrical cut). This is not surprising
because we know that for proton-proton the double pomeron exchange process
should be larger than the electromagnetic one for producing a light resonance.

In Table 4 we show the pion pair cross section for different ions. The
values were obtained using the geometrical cut, and even if with this procedure 
the diffractive
cross section is a little bit overestimated for heavy ions we verify that
photon-photon dominates. For light ions the diffractive process is already
of the order of $10 \%$ of the electromagnetic one.

The simulations discussed in the last paper of Ref.~\cite{klein} have shown
that the $\gamma\gamma$ interactions produce final states with small
summed transverse momentum $(| \sum \bar{p}_T |)$. Therefore, a cut
of $| \sum \bar{p}_T | \leq 40 \,\, - \,\, 100 \, MeV/c$ can reduce
considerably the background of non-peripheral collisions. 
In Table 4 we present the cross section for pion pair production
through double photon interaction with $| \sum \bar{p}_T | \leq 100 \,
MeV/c$. With this cut the cross section was reduced almost by a factor of $4$. 
The electromagnetic process with the restriction on
$p_T$ is still larger than those of double pomeron exchange without this 
cut, and the introduction of this cut in the diffractive process produces a
similar reduction.

The results for a hadron cluster production with invariant mass $M_X$ is
depicted in Fig.1. In the figure it is shown the cross section for
four different ions (Pb, Au, Ag, Ca) at energies that will be available
at RHIC and LHC. The results were obtained integrating the cross sections
with the condition $b_{min} > 2R$. At LHC the photon-photon process will
dominate the cross sections for heavy ions, whereas for light ions and
small invariant mass they become of the same order. For heavy ions
the diffractive process is indeed negligible. Note that our result for
photons is similar to the one of Engel {\sl et al.}~\cite{engel}, but
the diffractive cross sections is slightly smaller than the one of
Ref.~\cite{engel}. We credit this
deviation to the differences in our approachs to calculate the subprocess cross
section, mainly in the use of Eq.(\ref{l3}) with the changes prescribed by the
Donnachie and Landshoff~\cite{land2} instead of Eq.(\ref{sigmar}) given by the
Ter-Martirosyan \cite{mart} model. They also use a value for $\sigma _0$ that
is smaller than the one we considered here, which gives a smaller cut of the
central collisions. We believe that the use of Eq.(\ref{l3}) and the model of
Ref.~\cite{land2} is more appropriate for the full range of momenta. Actually,
diffractive models are plagued by uncertainties and the measurement of the
double pomeron exchange in heavy ion colliders will provide useful information
to distinguish between different models.

For multiple particle production we will not have the criteria
of low multiplicity to help us to select the truly peripheral
collisions, as well as it is far from clear if the cut in transverse
momentum will be very effective to select the $\gamma\gamma$ events.
However, we can separate the peripheral events on the basis of a clustering
in the central region of rapidity, although an extensive
and detailed simulation of the background processes will be 
necessary in order to set the precise interval of rapidity needed
to cut the inelastic nuclear collisions.

As verified by Drees, Ellis and Zeppenfeld~\cite{papa}, Eq.(\ref{fatf}) is a
reasonable approximation for the form factor obtained from a Fermi or
Woods-Saxon density distribution. However, their result shows that for heavy
final states the photon-photon luminosity is slightly underestimated, and we
can expect the same for the Pomeron one. A simple form factor expression
consistent with the Fermi distribution has been recently obtained in
Ref.~\cite{kn}, and its use would yield a few percent larger cross section in
the case of a very heavy hadron cluster production. 

In the case of peripheral heavy ion collisions at RHIC we surely cannot
neglect the diffractive contribution, for light ions and a hadron cluster 
with low invariant
mass it surely dominates photon-photon collisions. Notice that these
results may change if we use the Glauber absorption factor to compute
the cross section (depending on the energy, the ion and invariant mass),
but the actual fact is that double pomeron exchange cannot be neglected
at RHIC.

In conclusion, we estimated the production of resonances, pion pairs and
a cluster of hadrons with invariant mass $M_X$ in peripheral heavy ion
collisions at energies that will be available at RHIC and LHC. The
condition for a realistic peripheral collision was studied with the
use of a geometrical cut, where the minimum impact parameter
was forced to be larger than twice the identical nuclei radius. The introduction 
of an absorption factor in the Glauber approximation to eliminate central
collisions was also studied. We find out that the most restrictive method to
account for inelastic collisions depended on the energy, the ion, as well as
on the value of  $\sigma_0$ (the nucleon-nucleon total cross section). The
geometrical cut is not allways the most restrictive way to enforce peripheral
collisions, an this is a topic that should be answered by the future
experiments. In both cases we noticed that at energies of the LHC operating in
the heavy ion mode and for very heavy ions the double pomeron exchange is not
a background for the two photon process. The situation changes considerably
for light ions and mostly for the energies available at RHIC, where double
pomeron exchange cannot be neglected.   

\section*{Acknowledgments}
One of us (C.G.R.) thanks 
Paulo S. R. da Silva for useful discussions, and C. A. Bertulani for a helpful
remark.
This research was supported in part by 
the Conselho Nacional de Desenvolvimento Cientifico e Tecnologico (CNPq) (AAN),
Funda\c c\~ao de Amparo a Pesquisa do
Estado de S\~ao Paulo (FAPESP) (CGR and AAN), and by Programa de Apoio a
N\'ucleos de Excel\^encia (PRONEX).

\newpage


\begin{figure}[htb]
\epsfxsize=.9\textwidth
\begin{center}
\leavevmode
\epsfbox{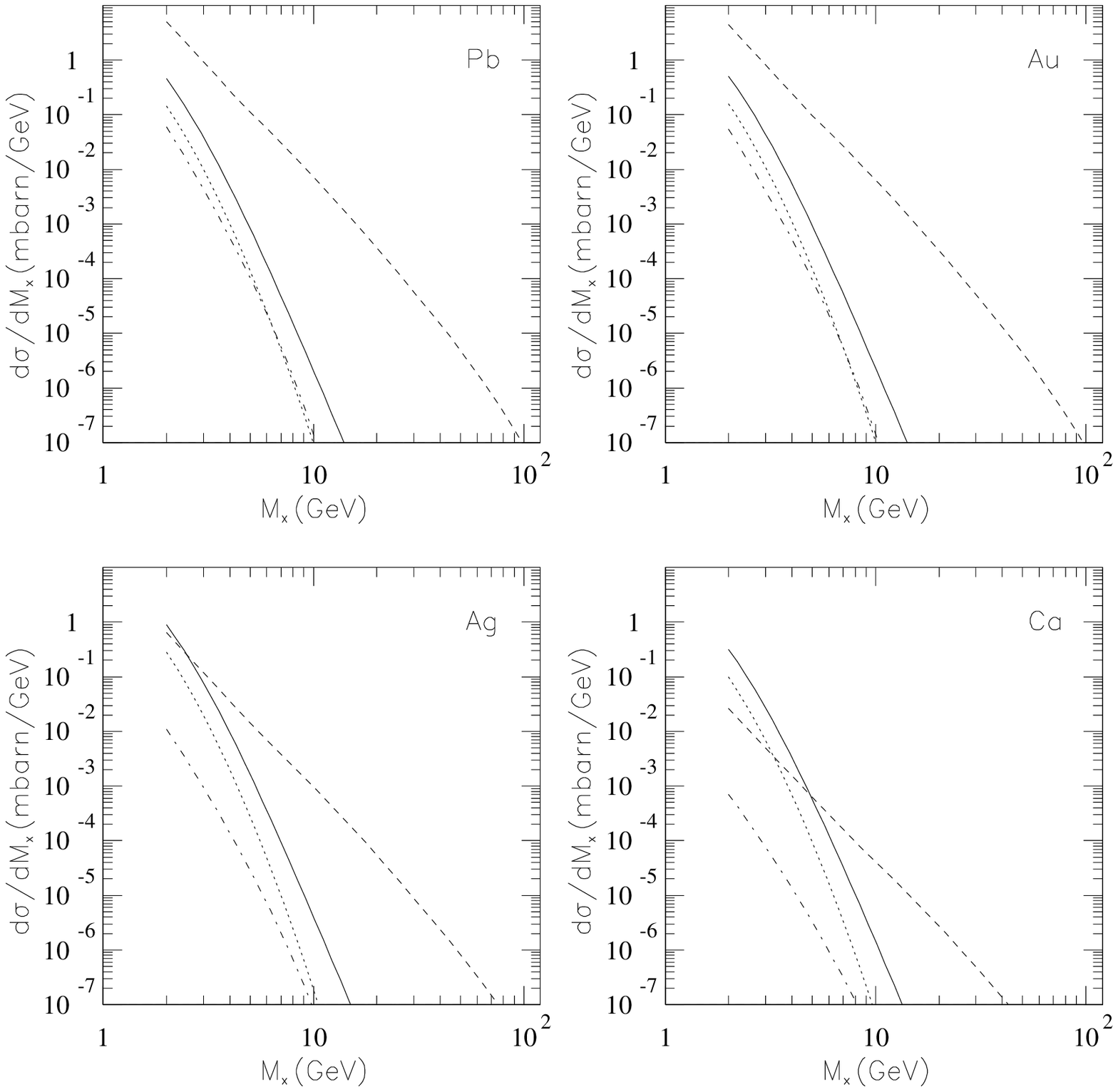}
\end{center}
\vskip -3.cm
\caption{ Cross section  for multiple particle production with invariant 
mass equal to $M_X$ for different nuclei collisions. The nuclei are indicated
in the upper corner of each figure. The solid line is for pomeron-pomeron
interaction and the dashed line is for double photon exchange at $LHC$,
$\sqrt{s} = 6 300$ GeV/nucleon. In the same figures it can be seen the cross
section  for $RHIC$, $\sqrt{s} = 200$ GeV/nucleon. Double pomeron exchange is
given by the dotted line and the photon interaction by the dotted-dashed
line.}  \label{ptxyphi} 
\end{figure}
\vskip .7cm


\begin{table}[hb]
\center
\begin{tabular}{l c c c c c c}
Meson & $M_R$ & $\Gamma_{(R \rightarrow \gamma \gamma)}$ & 
{\it RHIC}$_{\gamma \gamma}$ & {\it LHC}$_{\gamma \gamma}$ &
  {\it RHIC}$_{\cal PP}$ & {\it LHC}$_{\cal PP}$ \\ 
\hline
$\pi^0$ & 135  & $8 \times 10^{-3}$ &  7.1 & 40 &
0.05  & 0.367  \\
$\eta$ & 547  & 0.463 & 1.5 &17 &
 0.038 & 0.355   \\
$\eta ^\prime$ & 958  & 4.3 & 1.1 & 22  &
0.04  & 0.405 \\
$\eta_c$ &  2979 & 6.6 &$0.32 \times 10^{-2}$ & 0.5 & 
$0.47 \times 10^{-4}$ & $0.27 \times 10^{-3}$  \\
$\eta ^\prime _c$ & 3605  & 2.7  & $0.36 \times 10 ^{-3}$ & 0.1 &
 $0.34 \times 10^{-5}$ & $0.61 \times 10^{-4}$ \\
$\eta _b$  & 9366  & 0.4 & $0.13 \times 10^{-7}$ & $0.37 \times 10^{-3}$  &
$0.11 \times 10^{-10}$ & $0.77 \times 10^{-9}$  \\ 
\end{tabular}
\label{tab1}
\caption{Cross sections for resonance production through
photon-photon ($\gamma\gamma$) and double-pomeron ($PP$) processes. 
For $RHIC$, $\sqrt{s} = 200$ GeV/nucleon, we considered $^{238}$U ion and for
$LHC$, $ \sqrt{s} = 6 300$ GeV/nucleon, the nucleus is $^{206}$Pb. The
cross sections are in mbarn.} 
\end{table}

\vskip 1.0 cm

\begin{table}[hb]
\begin{center}
\begin{tabular}{l c c } 
Meson & $\sigma_{AA}^{gl} / \sigma^{{\cal PP} \rightarrow R}_{AA}$ ({\it
LHC}) & $ \sigma_{AA}^{gl} / \sigma^{{\cal PP} \rightarrow R}_{AA}$ ({\it
RHIC})     \\ \hline
$\pi^0$ & $3.54 \times 10^{-3}$ & $1.5 \times 10^{-2}$ \\
$\eta $ & $3.58 \times 10^{-3}$ &  $1.47 \times 10^{-2}$ \\
$\eta ^\prime$ & $3.46 \times 10^{-3}$  & $1.5 \times 10^{-2}$ \\
$\eta _c$ &$3.47 \times 10^{-3}$  & $1.32 \times 10^{-2}$ \\
$\eta ^\prime _c$ & $3.61 \times 10^{-3}$ & $1.5 \times 10^{-2} $\\
$\eta _b$ &$3.5 \times 10^{-3}$ & $1.45 \times 10^{-2}$ \\
\end{tabular}
\caption{ Ratios of cross sections for diffractive resonance production 
calculated with the Glauber absorption factor to the one with the geometrical
cut in
the collision of $^{238}$U for energies available 
at RHIC ($\sqrt{s} = 200 $ GeV/nucleon), and collisions of $^{206}$Pb for
energies available at LHC ($\sqrt{s} = 6.300$ GeV/nucleon).}  
\end{center} 
\label{tab5}
\end{table}
\begin{table}[hb]
\begin{center}
\begin{tabular}{l c c c c } 
Nucleus & $\sqrt{s}$ & $\sigma ^{{\cal PP} \rightarrow R}_{AA}$  &
$\sigma^{gl^{\cal PP}}_{AA}$ & $\sigma _{\gamma \gamma }$   \\
\hline
Au ($A$=197) & 100  & 0.044 & $0.55 \times 10 ^{-3}$ &  2.4 \\ 
Ca ($A$=40) & 3 500 & 0.043 & $0.39 \times 10^{-3}$ & 0.14 \\
Si ($A$=28) & 200 & $0.34 \times 10^{-2}$ & $0.15 \times 10^{-3}$ & $0.69 \times 10^{-2}$ \\
Si ($A$=28) & 100 & $0.22 \times 10^{-2}$  & $0.12 \times 10^{-3}$ & $0.39 \times 10^{-2}$ \\
\end{tabular}
\caption{Cross section for $\pi^0$ production for
different ions and at different energies. The energies are in GeV/nucleon and
the cross sections in mbarn. $\sigma ^{PP \rightarrow R}$ is the cross section computed 
with the geometrical cut and $\sigma ^{gl}$ is the one with the absorption
factor.}   
\end{center}
\label{tab3}
\end{table}
%
\begin{table}[hb]
\begin{center}
\begin{tabular}{l c c c c}
Nucleus & $\sqrt{s}$ & $\sigma_{\gamma \gamma}$ & $\sigma _{\gamma \gamma}
(p_T < 100 \, \mbox{MeV})$ & $\sigma_{\cal PP}$ \\  \hline
U  & 200 & 9 & 2.15 &  $7.47 \times 10^{-3}$ \\
Pb & 6 300 & 81.96 & 15.98 & $1.34 \times 10^{-2}$  \\
Au & 100 & 2.3 & 0.523 & $8.11 \times 10^{-3}$   \\
Ca & 3 500 & 0.28 & 0.05 & $5.85 \times 10^{-3}$  \\
Si & 200   & $0.98 \times 10^{-2}$ & $0.21 \times 10^{-2}$ &$1.02 \times 10^{-3}$    \\
Si & 100   & $0.49 \times 10^{-2}$ & $0.12 \times 10^{-2}$  &  $8.6 \times 10^{-4}$  \\  
\end{tabular}
\caption{Cross sections for $\pi^+ \pi^-$ production. 
The energies are in GeV/nucleon  and the cross sections in mbarn. 
$\sigma _{\gamma \gamma} (p_T < 100 \, \mbox{MeV})$  is the pion pair
production through photon-photon interaction with the cut $p_T
<100$ MeV  .}   
\end{center} \label{tab6}
\end{table}

\end{document}